\documentclass[reqno, 11pt]{article}% Math and Physical Sciences Numbered Reference Style 

\usepackage{graphicx,tikz}%
\usepackage{multirow}%
\usepackage{amsmath,amssymb,amsfonts}%
\usepackage{amsthm}%
\usepackage{mathrsfs}%
\usepackage[title]{appendix}%
\usepackage{xcolor}%
\usepackage{textcomp}%
\usepackage{manyfoot}%
\usepackage{booktabs}%
\usepackage{algorithm}%
\usepackage{algorithmicx}%
\usepackage{algpseudocode}%
\usepackage{listings}%
\usepackage{chemformula}
\usepackage{siunitx}
\usepackage{authblk}
\usepackage{amssymb,latexsym,amsmath,amsfonts}
\usepackage{graphicx}
\usepackage{color,enumerate}
\usepackage{amsmath,amsfonts,amssymb}
\usepackage[mathscr]{eucal}
\usepackage{bbm,pgfkeys}
\usepackage{booktabs}
\usepackage{float,xcolor}
\usepackage{hyperref}
\usepackage{amsthm}
\usepackage[mathscr]{eucal}
\usepackage{paralist}
\usepackage{rotating}
\usepackage{multirow}
\usepackage[left=1in, right=1in,bottom=1in,top=1in]{geometry}
\hypersetup{
	colorlinks=true,         % Enables colored links (no boxes)
	linkcolor=red,           % Color for internal links (e.g., figures, equations)
	citecolor=blue,          % Color for citations
	urlcolor=blue,           % Color for URLs
}

\usepackage{adjustbox}

\definecolor{medgreen}{rgb}{0, 0.75, 0}
\definecolor{darkgreen}{rgb}{0, 0.3, 0}

\theoremstyle{definition}

\theoremstyle{remark}

\newcommand{\RN}[1]{\uppercase\expandafter{\romannumeral#1}}

\title{Towards a unified framework for multiple stable states in ecological systems}

\author[1]{Jennifer Paige}
\affil[1]{Department of Mathematics, University of California, Davis, USA}

\author[2]{Denis D. Patterson\thanks{Corresponding author: denis.d.patterson@durham.ac.uk}}
\affil[2]{Department of Mathematical Sciences, Durham University, UK}

\author[3,4]{Alan Hastings}
\affil[3]{Department of Environmental Science and Policy, University of California, Davis, USA}
\affil[4]{Santa Fe Institute, Santa Fe, USA}

\date{\today}

\begin{document}

\maketitle

\begin{abstract}
Multiple stable states--the coexistence of two or more distinct ecological configurations under identical environmental conditions--have attracted sustained interest in ecology, yet the field still lacks a unified framework connecting ecological mechanisms to dynamical models. Here, we review empirical and theoretical approaches to multiple stable states, synthesising perspectives on stability, tipping, hysteresis, and transient dynamics, and contextualise these within a common mathematical framework. Drawing on examples of well-known ecosystem models, we highlight the central and necessary role of positive feedback loops and identify other common, unifying features of ecological systems that exhibit multiple stable states. We further discuss the relationship between stable and transient dynamics, the roles of spatial and temporal scales in feedback identification, and the implications for ecological restoration and management. We conclude with open questions and challenges for the field, including extending multistability theory to persistent-transient frameworks and harnessing emerging data-collection technologies to sharpen empirical inference.
\end{abstract}

\section{Introduction}

Multiple stable states--frequently referred to as alternative stable states in the case of two stable outcomes--have long been a topic of interest for empiricists and theoreticians alike~\cite{watt1947pattern}. If, after a disturbance, an ecological system does not return to its previous state, does this signal permanent change, or might multiple system outcomes be possible? If in a system with coral and algae, different patch reefs are found to have significantly different proportions of coral cover~\cite{mumby2013evidence}, does this necessarily mean that the patch-level environmental conditions are different, or are different ecological configurations possible for these reefs? Crucially, multiple stable states in ecological systems can explain why different species exhibit different abundances under the same physical conditions.

Numerous observations of ecological systems suggest the possibility of alternative outcomes~\cite{petraitis2013multiple,petraitis1999experimental}. However, debate continues over which systems definitively exhibit multiple stable states~\cite{mumby2007thresholds,mumby2013evidence}.  What empirical evidence constitutes sufficient proof of the existence of multiple stable states? When can collecting data on a large spatial scale be a useful proxy for long-term time observations? Notable examples of well-studied empirical systems claimed to exhibit multiple stable states include shallow lakes~\cite{scheffer1989alternative,scheffer1993alternative}, coral reefs~\cite{mumby2007thresholds,mumby2013evidence,schmitt2019experimental}, and tropical forest-savanna ecosystems~\cite{staver2011global,staver2011tree}. For these three examples, there are also well-studied mathematical models that illustrate the mechanisms leading to multistability~\cite{scheffer1989alternative, mumby2007thresholds, staver2011tree}. Given the observation of what appear to be similar dynamical patterns in these different systems, what do these systems, and possibly the models developed for them, have in common?  Is there a similar underlying mechanism generating multiple stable states in these systems?

From a modelling standpoint, even simple ecological models, such as two-species Lotka-Volterra competition models or single-species models with the Allee effect, can have multiple stable equilibria~\cite{courchamp2008allee, volterra1928variations}. One could even think of the classic Levins patch model as modelling each patch switching between two stable states: occupied or unoccupied~\cite{levins1969demographic}. Thus, simple models demonstrate that multiple stable states are not difficult to achieve in theory. However, these simple models are not fully mechanistic, but rather are phenomenological, and this begs the question: are there particular kinds of ecological interactions that guarantee either that a system must have or cannot have multiple stable states? More broadly, is a focus on the long-term dynamics of ecological models the right way to explain observations of alternate states in real-world systems? Perhaps most pressing of all: what are the implications for ecosystem management?

To answer these questions, we first need a working definition of multiple stable states. Providing a precise definition of multiple stable states in an ecological system is not straightforward, as spatial and temporal aspects play important roles. Additionally, while definitions of terms for modelling multiple stable states can be easy to establish, demonstrating these same definitions in observational studies may be infeasible. Not surprisingly, given that the word stable is in the name of the phenomenon, concepts of stability play an important role in describing and understanding multiple stable states. Moreover, multiple stable states are only interesting if systems can naturally be found in more than one state and, importantly, can transition between them in response to perturbations; here, the notion that stability is only defined relative to the class of relevant perturbations is key to unpacking its role in multiple stable states~\cite{lewontin1969}. Stability is often taken in the mathematical sense to refer to local asymptotic stability, meaning only stable with respect to arbitrarily small perturbations--thus, we are looking at systems where sufficiently large perturbations change the system, so it does not return to the previous stable state. These perturbations can be changes to the state of the system (species populations) or to the conditions governing the system (model parameters), although what is considered a parameter or a state variable is often a modelling choice. This distinction maps naturally onto two important classes of tipping~\cite{scheffer2001catastrophic,hastings2026tipping}, bifurcation-induced tipping (B-tipping) and noise-induced tipping (N-tipping), both of which are central to understanding the dynamic implications of multiple stable states and the potential for regime shifts in a multistable system.

Beyond tipping~\cite{scheffer2001catastrophic}, concepts such as hysteresis~\cite{blackwood2012effect,staal2020hysteresis}, thresholds~\cite{may1977thresholds,mumby2007thresholds}, catastrophe theory~\cite{ludwig1978}, transients~\cite{hastings2018transient}, positive feedback~\cite{deangelis2012positive,suding2004alternative} and the effect of spatial scale~\cite{levin1992problem} are often associated with systems that have multiple stable states. Exploring the underlying dynamics of an ecological system reveals a complicated picture of nonlinear responses and dependencies. As conditions change, understanding exactly when and how dramatic any resulting change in dynamics can be is challenging~\cite{scheffer2009early,boettiger2012quantifying}. Empirical data collection and analysis to anticipate a significant shift or tipping point in dynamics, also known as early warning signals, are still in development~\cite{scheffer2009early, boettiger2012quantifying, boettiger2013early, ditlevsen2023warning}. Predicting and acting before these significant and difficult to reverse changes occur is an important factor in management decisions~\cite{hastings2016timescales}. Especially in a highly disturbed system, understanding and either buffering or restoring in an efficient way is critical~\cite{maes2024explore,suding2004alternative}.

In Section \ref{sec.previous}, we survey previous studies to explore the common ecological features that lead to multiple stable states. We also review how empirical data has been used to argue for and against ecosystems exhibiting alternative stable states, highlighting the key examples of shallow lakes, coral reefs and forest-savanna systems. In Section \ref{sec.math}, we discuss and contextualise the definitions of the relevant mathematical concepts and emergent properties that unify the example systems discussed, with a particular focus on the definition and role of positive feedback loops. Finally, in Section \ref{sec.discussion}, we discuss the implications of our synthesis for the field and highlight the most pressing challenges and opportunities for advancing a unified theory of multistability in ecology.

\section{Multiple Stable States in Ecology}\label{sec.previous}

\subsection{Historical development}\label{sec.history} 
As the idea of ecological systems having multiple stable states has developed, several recurring themes emerge: the debate of empirical criteria of multiple stable states, the role of scale in defining a landscape or patch or in defining ecologically relevant timescales, the formalisation of concepts of stability, and feedback as a possible underlying mechanism. Across this body of work, the question of how widespread multistability is, and what underlying commonalities the systems share, have remained persistent and motivating themes.

A.S. Watt, through his synthesis of seven plant communities in 1947, was a pioneer in raising the idea that an ecosystem can present multiple different configurations despite a similar environment, pushing the ecological description of patterns and mechanisms beyond the individual and into the community-level~\cite{watt1947pattern}. On this idea of underlying mechanisms, Watt noted that a feedback process governed the turnover of community composition--with communities building up (``upgrading'') and breaking down (``downgrading''). Patchiness of a landscape seemed an important clue to Watt of the existence of multiple equilibria and its possible ``widespread'' occurrence, and he began exploring the idea that the proportion of a landscape occupied by each phase could serve as an indicator of its dominance in the community cycle.

Observation of more ecological systems that exhibited characteristics consistent with multiple outcomes began to emerge, such as Sutherland's 1974 study of fouling communities--which led him to conclude that ``multiple stable points are an undeniable reality''~\cite{sutherland1974multiple}. However, Sutherland was unable to replicate his results and while a useful indicator, Watt's observation of ``patchiness'' was seen as insufficient empirical evidence for the existence of multiple stable states by modern standards. Rigorously proving that an ecological system has multiple stable states became a point of controversy, with several authors developing diverging criteria for proof~\cite{connell1983evidence,peterson1984rigorous}. Arguments against emerging evidence of multiple stable states often criticised insufficient data, particularly in proving environmental conditions were uniform across different patches or in proving that no external forces artificially influenced the dynamics~\cite{connell1983evidence}.

Core to this debate were differing perspectives on appropriate scales (both spatial and temporal) and on definitions of stability. On scale, Sutherland rejected the notion that a more appropriate spatial scale could erase the emergent historical contingency, but did question the proper choice to define community states and stability, ultimately deferring to ``naturalist's judgment''~\cite{sutherland1974multiple}. Later, scientists argued for more stringent scales, focused specifically on the smallest scale of ecological and community relevance~\cite{connell1983evidence, peterson1984rigorous}. Levin, in his 1992 MacArthur lecture, argued that there is no one correct choice in a selection of scale, leaving empirical scientists and ecological modellers a decision that could impact what is captured from the system~\cite{levin1992problem}. On stability, Lewontin, in 1969, proposed the adoption of advanced mathematical techniques--specifically vector field notation and definitions (e.g. ``classical theory of neighbourhood stability'')--to formalise analysis of ecological systems, creating a framework to better describe the dynamical underpinnings, such as identifying stationary, transient, and cyclical points~\cite{lewontin1969}. May drew on this framework, that complex dynamical systems can be decomposed into lower-dimensional interactions, to highlight the connection between empirical evidence and the mathematical understanding of multiple stable states, emphasising a more unified modelling framework in several ecological examples, such as grazing communities, animal harvesting, insect pests, and human host-parasite systems~\cite{may1977thresholds}. However, this definition of stability was not uniformly adopted in the literature, and others deferred to definitions based on long-term persistence, resilience to stochasticity, or resistance to change~\cite{sutherland1974multiple, connell1983evidence, petraitis2013multiple}. Even within this mathematically formalised framework, distinguishing the class of perturbations that make a state stable with ecological relevance remained debated~\cite{connell1983evidence, petraitis2013multiple}.

The special consideration of the role of feedbacks developed further in the literature from Watt's initial paper, although the terminology and definitions used differed throughout~\cite{watt1947pattern, sutherland1974multiple, connell1983evidence, peterson1984rigorous}. Specifically, an influential 1992 text by Wilson and Agnew focused on positive feedback loops, or ``switches,'' in which a community modifies the environment to make it more suitable for that community (i.e., ecosystem engineering)~\cite{wilson1992positive}. They directly tied these switches to the formation of stable mosaics--sharp boundaries between vegetation types along an environmental gradient--and the acceleration or delay of succession. On the other hand, Petraitis in his more recent, comprehensive book on multiple stable states, took a different definition of positive feedback (potentially implying it must consist of positive interactions), to directly critique Wilson and Agnew's focus on positive feedback loops, or more generally, any common driving characteristic of multiple stable states. He instead argued that the more interesting question is how frequently such characteristics arise in systems that do have alternative states, not as a requirement, but as a common overlap~\cite{petraitis2013multiple}. 

\subsection{Empirical evidence}
Empirically, characterising and understanding systems with multiple stable states is challenging and requires careful interpretation of field data. Similarly, translating ideas between the empirical and theoretical study of multiple stable states is not straightforward, although significant progress has been made in recent years~\cite{beisner2003alternative,scheffer2001catastrophic}. While the mathematical definition of stability in model outcomes is well-defined (see Section~\ref{sec.math_prelim}), ensuring that this stability reflects biological reality is often more nuanced and less clear-cut. Furthermore, identifying the ecological mechanisms that drive stability, or the mechanisms underlying a possible switch to an alternative stable state, requires careful investigation. Observation can yield insights into relationships among components (biotic or abiotic) of the ecosystem, and observing a regime shift in a system can provide clues in identifying key drivers or possible stable configurations. However, such analysis does not necessarily establish the existence or cause of multiple stable states~\cite{peterson1984rigorous}.

\begin{figure}
    \centering
    \includegraphics[width=0.99\linewidth]{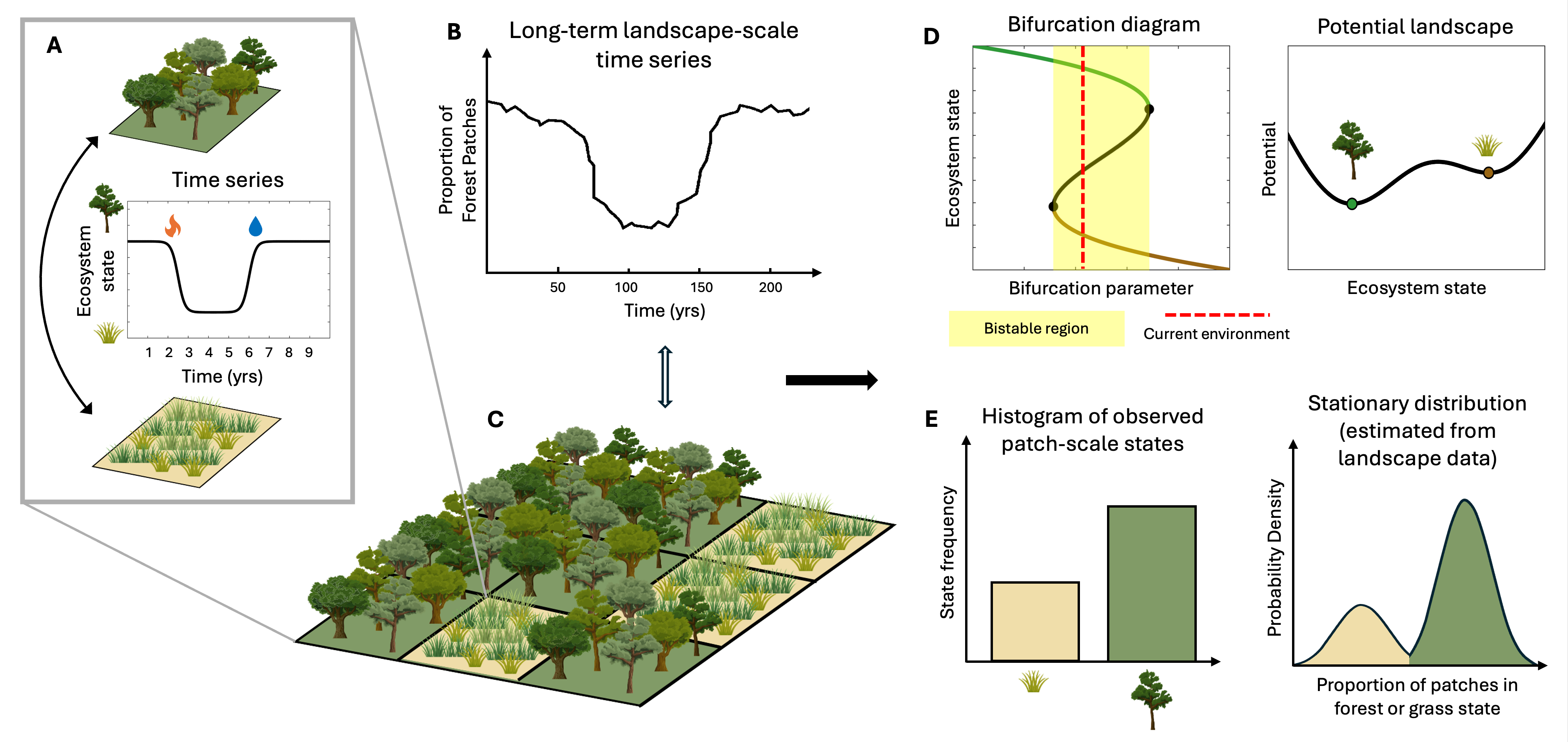}
    \caption{Illustration of the space-for-time substitution in ecosystems with multiple stable states. \textbf{A}: Time series of ecological state for a single patch, switching between two stable states over time. \textbf{B}: Long-term, landscape-scale time series showing the aggregate proportion of patches in each state, reflecting the overall system dynamics. \textbf{C}: A spatial snapshot of the landscape at a single moment in time, composed of many patches each occupying one of the two stable states; the space-for-time substitution treats this snapshot as a proxy for the temporal dynamics shown in \textbf{B}. \textbf{D}: Bifurcation diagram and potential landscape for the system, illustrating the bistable parameter regime within which both stable states coexist. \textbf{E}: The histogram of observed patch-scale states across the landscape (left) and the corresponding bimodal stationary distribution inferred from landscape-level data (right), whose two modes correspond to the two stable states identified in \textbf{D}.}
    \label{fig:space_for_time}
\end{figure}

One approach to establishing the existence of multiple stable states is to conduct a controlled experiment in which all environmental variables are held constant, demonstrating that different initial population sizes lead to convergence on distinct, persistent states~\cite{connell1983evidence, peterson1984rigorous}. Demonstrating hysteresis (see Section~\ref{sec.math_prelim}) requires additionally showing that the pathway back to a stable state differs from the pathway away from it. This distinction maps onto the difference between pulse perturbations, which are transient disturbances that leave underlying conditions unchanged, and press perturbations, which impose a sustained shift in a system parameter, with the latter more informative for detecting hysteresis~\cite{petraitis2013multiple}. In practice, the aforementioned types of experiments are largely infeasible given the spatial scale of the systems of interest, as well as conservation and ethical constraints. More limited experimental designs and observational proxies are therefore typically used to assess the likelihood of multiple stable states. These include the ``space-for-time'' substitution, long-term records documenting abrupt and asymmetric transitions or paleoecological archives that extend the temporal window of observation (see Figure~\ref{fig:space_for_time}B), and manipulative mesocosm experiments. Each approach carries its own assumptions and limitations, raising nuanced mathematical and statistical questions about what conclusions can be drawn~\cite{petersen2001dimensional,petraitis2013multiple}.

The space-for-time approach treats the observed variation in ecological state across spatially distinct patches (sharing the same general environmental conditions) as a proxy for the long-run temporal dynamics of a single patch (see Figure~\ref{fig:space_for_time}A-C); this idea has roots in Watt's original analysis of patch dynamics~\cite{watt1947pattern}. The space-for-time substitution is especially valuable in ecology because the full signature of multiple stable state dynamics rarely becomes apparent over timescales amenable to observation. A partial mathematical justification for this approach can be drawn from ergodic theory: for a dynamical system that is ergodic with respect to an invariant measure, time averages converge to spatial averages, so that a cross-sectional snapshot of many patches is equivalent to observing a single patch over a long time horizon~\cite{walters2000introduction}. However, this ergodic justification is limited as trajectories initialised in distinct basins of attraction remain confined to those basins and do not explore the full state space, violating the ergodicity assumption. The substitution is therefore better understood as resting on the weaker requirement that the ensemble of observed patches samples multiple basins of attraction, and that each patch has had sufficient time to approach its local attractor (Figure~\ref{fig:space_for_time}D). Under these conditions, bistability manifests in the spatial snapshot as a bimodal distribution of observed patch states, with modes corresponding to the two attractors and a trough near the unstable equilibrium separating them (see Figure~\ref{fig:space_for_time}E). In practice, neither condition is guaranteed. Ecological systems are rarely observed in uniform environments, and increasing disturbance frequency or intensity may further complicate the detection and analysis of systems with multiple stable states. Moreover, the states observed may not be truly stable, but instead persistent transients. All of these complications confound simplifying assumptions that permit straightforward space-for-time substitution. When temporal environmental change exceeds the range of spatial variability at a moment in time, significant discrepancies may arise, and long-term observation is still most appropriate for full system understanding~\cite{blois2013space, kreyling2025spacefortime}.

Shallow lakes are one of the most well-studied ecosystems proposed to exhibit multiple stable states~\cite{scheffer1993alternative}. These systems are theorised to exist in either a clear-water state, dominated by submerged macrophytes, or a turbid, phytoplankton-dominated state, with a self-reinforcing feedback maintaining each: macrophytes stabilise water clarity by reducing sediment resuspension and suppressing phytoplankton, while dense phytoplankton blooms shade out macrophytes and perpetuate turbidity. Scheffer formalised these mechanisms in simple nonlinear models, demonstrating hysteresis and the potential for abrupt transitions between states as nutrient levels cross critical thresholds~\cite{scheffer1989alternative}. Subsequent empirical work across northern Europe and North America generally supports this theoretical framework. However, debate persists, with some authors attributing observed regime shifts to transient dynamics or external disturbances~\cite{carpenter2003regime,jeppesen2005lake}, and others pointing to hysteresis in recovery trajectories following nutrient reduction as evidence of multiple stable states~\cite{ibelings2007resilience}.

Coral reefs are another notable example of an ecosystem proposed to have multiple stable states: a healthy reef with high coral cover and a degraded reef, dominated by macroalgae. Grazing fish populations mediate the transition between these two states, reducing algal cover on the reef and maintaining a healthy coral population. High macroalgae populations can overcome moderate grazing and even benefit from it, thereby facilitating further colonisation of coral and increasing coral mortality. Alternatively, high coral cover allows the newly created space opened by grazing to be dominated by coral settlement, thereby maintaining coral dominance. These mechanisms were elucidated in a simple phenomenological model and shown to facilitate multiple stable states by Mumby et al.~\cite{mumby2007thresholds}. Critique emerged, surrounding the lack of proof on a global scale of reefs ``tipping'' into macroalgae-dominant states as proposed and about the specific choice of a hysteretic curve to describe the state change~\cite{bruno2009assessing, dudgeon2010phase}. Mumby et al. responded with an acknowledgement of the challenges to experimental evidence that a system with slow dynamics and a complex disturbance regime can present~\cite{mumby2013evidence}. This debate shows the difficulties of applying the space-for-time argument in practice, as reefs are unlikely to be in ``true'' equilibrium.

In the tropics, it has been proposed that open-canopy savanna and closed-canopy forest are alternative stable states~\cite{staver2011global}. Satellite-derived tree cover data, combined with mean annual rainfall (MAR) records, show that tree cover exhibits a distinctly bimodal distribution in regions with intermediate MAR. Supported by mathematical models, the most widely accepted explanation for the observed bimodal distribution of forests and savannas is that regular disturbance by fires can maintain open-canopy savannas~\cite{staver2011tree,staver2012integrating}. In contrast, denser closed-canopy forests sufficiently suppress the flammable grass layer to limit fire spread and stabilise the forest state. This interpretation has been challenged on the grounds that environmental heterogeneity, rather than true bistability, may explain the observed patterns~\cite{wuyts2017amazonian,higgins2024reassessing}, though recent work suggests bistability may actually be underestimated once climate-vegetation feedbacks are accounted for~\cite{staal2020hysteresis}.

Across these three systems, common threads emerge from both the theory and the debate: self-reinforcing interactions that maintain each state, persistent questions about whether observed patterns reflect bistability, transient dynamics, or environmental variation, and the challenge of distinguishing genuine hysteresis from external forcing; themes that, as we shall see in Section \ref{sec.positive_feedback}, find a natural common language in the mathematical structure of the models proposed for each system. Table \ref{table:ecosystems} below surveys a broader range of ecological systems proposed to exhibit multiple stable states, summarising in each case the alternative configurations identified and the key mechanisms or processes thought to generate or maintain them.

\begin{table}[!htbp] 
\centering
%\begin{tabular}{p{2cm}p{5cm}p{4.5cm}p{2.5cm}}
\begin{tabular}{>{\raggedright\arraybackslash}p{2cm}>{\raggedright\arraybackslash}p{5cm}>{\raggedright\arraybackslash}p{4.5cm}>{\raggedright\arraybackslash}p{2.5cm}}
\textbf{Ecosystem} & \textbf{Alternative states} & \textbf{Mechanism(s)} & \textbf{References} \\
\hline
Coral reefs & coral‑dominated reef,  macroalgal‑dominated reef& herbivory, nutrient loading, physical disturbance &~\cite{mumby2007thresholds,blackwood2012effect,schmitt2019experimental}\\
\hline
 Kelp forest-urchin barren& kelp forest dominant reef (macroalgae dominant), urchin and crustose coralline algae dominant reef& recruitment facilitation, herbivory&~\cite{baskett2010recruitment, ling2015global}\\
\hline
Forest-savanna & closed-canopy forest, open-canopy savanna & fire, herbivory &~\cite{staver2011global,hirota2011global,aleman2020floristic}\\
\hline
Shallow lakes & turbid (algae dominated), clear & nutrient loading, macrophyte–water clarity feedbacks &~\cite{scheffer1997dominance,carpenter1999management,gilarranz2022regime}\\
\hline
Dryland ecosystems & vegetated (grass/shrub) state, bare‑soil/biocrust state & rainfall–infiltration feedback, local facilitation &~\cite{kefi2007local,chen2020biocrust,rietkerk2021evasion}\\
 \hline
%Infectious diseases & 
%low‑prevalence/disease‑free, high‑prevalence/endemic
% & re/super-infection, contact structure, immune waning/thresholds &~\cite{conway2015post,earn2000simple,gumel2012causes,nielsen2010bistable}\\ 
%\hline
Microbial ecosystems & alternative community compositions (e.g. different dominant species) & priority effects, mutualism, horizontal gene transfer, metabolic flexibility &~\cite{dubinkina2019multistability,amor2020transient,khazaei2020metabolic}\\
\hline
 Mussel beds& developed mussel bed, absence of mussels (sediment covered barren or rockweed cover)& clarity feedbacks (environment alteration, ecosystem engineering)&~\cite{coco2006feedbacks, petraitis1999experimental, petraitis2009experimental, petraitis2015variation}\\
%\hline
%Oyster beds?& well-developed high oyster reefs, sediment-covered, degraded low oyster reefs
%(oyster bed alternative stable states lit also includes oysters/pisasters)& Clarity feedbacks (facilitated by oysters), recruitment feedbacks, or (possibly problematically) height of reef&~\cite{jordancooley2011bistability}\\
\end{tabular}
\caption{Examples of natural ecosystems suggested to exhibit or have the potential for multiple stable states. We also note the alternative states identified in the literature and the key mechanisms or physical processes proposed to generate or maintain these states in each ecosystem.}\label{table:ecosystems}
\end{table}
 
Systems with multiple stable states have important implications for management and restoration, precisely because their defining characteristics (nonlinearity, hysteresis, and tipping) mean that simply reversing the environmental change that caused a transition may be insufficient to restore a previous state. Loss of species diversity or key interactions can further erode the feedbacks that originally maintained a desired state, compounding the challenge of recovery~\cite{folke2004regime, suding2009threshold}. This suggests that effective restoration should target the feedback structure of the system directly, either by disrupting feedbacks that maintain a degraded state or by reinforcing those that support the desired one, and that combining biological and abiotic interventions is often necessary to address multiple constraints simultaneously~\cite{suding2004alternative, byers2006using}. Proactively preventing tipping is generally far less costly than attempting recovery after a transition has occurred, though identifying critical thresholds in advance remains difficult~\cite{scheffer2001catastrophic, scheffer2009early}. Early warning signals such as increased variance, critical slowing of dynamics, and flickering between states are promising tools for anticipating tipping points, but remain an active area of research~\cite{xu2023non}. Feedbacks and the tipping points they generate can operate across large spatial and temporal scales, adding further practical challenges to restoration; restoring metapopulation connectivity and accounting for historical contingency, such as the role of seed banks, may be essential to re-establishing the interactions that sustain a recovered state~\cite{levin1992problem, suding2004alternative}.

\subsection{Theoretical approaches} 
More recently, the field has moved to more strongly contextualise the topic of multiple stable states within other ideas in the broader literature. 
%metapopulations
For example, essential to the idea of multiple stable states is the idea of landscapes having subunits that may display distinct dynamics. The concept of a metacommunity, or a connected group of local-scale ecosystem patches composing a broader landscape, dates back to Levins~\cite{levins1969demographic}. In the context of multiple stable states, metacommunities may impact patch dynamics via dispersal or mixing, creating the opportunity for either population sources or sinks depending on the configuration and patch states~\cite{leibold2004metacommunity, vandeleemput2015resilience}. As found by Van De Leemput et al., these connections can stabilise a community and increase the tolerance to disturbance before transitioning states~\cite{vandeleemput2015resilience}. This idea of multiple patches interacting further reinforces the importance of scale in fully understanding a system.
%Could also bring up edge effects -- impacts of imposing boundaries around sub-units
%Wilson, D.S. (1992). Complex interactions in metacommunities, with implications for biodiversity and higher levels of selection. Ecology, 73: 1984-2000

%tipping and catastrophe
Multiple stable states also serve as an important example in topics such as tipping and catastrophe theory. In a foundational paper, Scheffer et al. connect ecological systems of this form to the broader pattern that even small environmental changes can lead to catastrophic shifts~\cite{scheffer2001catastrophic}. This explicit tie to catastrophe theory aids in framing how the ecosystems transition between states, specifically highlighting the nonlinearity and abruptness of the transition. In multistable systems, these transitions can be caused by a reduction in the basin of attraction, tied to the environmental change of a desired ecosystem state, or noise in the system pushing the community out of the basin of attraction~\cite{hastings2026tipping}. Importantly, while systems with multiple stable states are good examples of these topics, it is not necessary to have multistability to display tipping or catastrophic change.
%It is worth noting that not all parameter changes causing a regime shift necessitate the ecosystem have multiple stable states, and both states must be achieved in the same environmental (parameter) states to reach multi-stability.
%By reviewing evidence from lakes, coral reefs, oceans, forests, and arid lands, they conclude that a loss of resilience typically precedes abrupt regime regime shifts, which are frequently governed by hysteresis. 
%Scheffer et al. highlight that this switch between states can be quite difficult to predict, with insufficient early warning signals. 

%stability
%affects whether both states are always present in analysis and also affect the understanding of resilience of a system, a frequently used but quite broad term
%transients
Multiple stable states have also been studied from many perspectives within theoretical ecology, creating diversity in understandings of stability. As addressed by Beisner et al., alternate perspectives, arising from a community assembly framework or an ecosystem environmental dynamics framework, lead to the framing of multiple stable states as arising from population (initial condition) disturbance or alternately environmental (parameter) change~\cite{beisner2003alternative}. These differing perspectives ultimately formalise in modelling choices (parameters vs. variables). Fukami et al., using a community assembly analysis, even argue that focusing on true stability in analysing systems of multiple stable states may overlook or misrepresent important ties to transient dynamics. They argue that underlying the stable state analysis are the assumptions that the long-term stable dynamics well represent short-term ecosystem trends and non-conforming transient dynamics do not persist on relevant timescales, which they conclude may not always hold in ecological systems~\cite{fukami2011commmunity}. Given that definitive proof for multiple stable states (especially in contrast to multiple transient states) may be infeasible to collect, a strong argument arises for explicit consideration of and connections to transient dynamics~\cite{fukami2011commmunity}.

\section{Unifying Features} \label{sec.math} %(3000 words)}

Building on this established understanding of previous studies of multiple stable states and related topics, we now synthesise previous work into a new, cohesive, mathematical framework. We begin with a careful discussion of stability and associated concepts, before turning to mechanisms, presented both biologically and mathematically, and with special attention to the core concept of a positive feedback loop. These approaches are then illustrated through several key examples of well-known multistable ecosystem models. Through this exposition and analysis, a unifying picture emerges: positive feedback loops, made mathematically precise via the Jacobian sign structure, are a necessary feature common to systems with multiple stable states, and the loop diagram provides a convenient tool for identifying this structure in simple mathematical models.

\subsection{Mathematical Tools}\label{sec.math_prelim}
%There is often a significant translation gap between empirical and theoretical approaches, which can be exacerbated by differences in terminology and conceptualization. Overcoming this gap to better understand similarities between studied systems can reveal more broad conclusions and patterns than each example can illuminate. Better understanding when empirical observations of a system that likely possesses multiple stable states correspond well with the conceptualization of the system mathematically might reveal underlying system motifs. 
Before attempting a unifying analysis of multistable systems in ecology, we must first define the relevant terms and scope. Table~\ref{table:terms} summarises the key terminology used throughout this section, and we discuss each concept in more detail below. First, consider autonomous ordinary differential equation models of the form:
\begin{equation}\label{eq.ODE_model}
 \frac{d}{dt}\mathbf{x}(t) = f(\mathbf{x}(t)), \quad t \geq 0, \quad \mathbf{x}(0) \in \mathbb{R}^N_+,
\end{equation}
where $\mathbf{x}_i(t) \geq 0$ represents the abundance of species $i \in \{1,\dots,N\}$. Our use of the term ``species'' here is informal, and for many models, the term ``functional types'' would be more appropriate, especially given that many of the models we consider mix biotic and abiotic factors. Models in this class are ubiquitous in mathematical ecology and often serve as a first step in modelling approaches or hierarchies that subsequently incorporate additional features and complexities, such as stochasticity or spatial extent. ``Simple'' models, such as ~\eqref{eq.ODE_model}, highlight the key interactions that generate multiple stable states and serve as building blocks for more complex models built on the same underlying interactions and processes. We will also restrict ourselves primarily to the discussion of equilibrium or steady-state solutions to ~\eqref{eq.ODE_model}, i.e. $\bar{\mathbf{x}}$ such that $f(\bar{\mathbf{x}}) = \mathbf{0}$, for ease of exposition and mathematical tractability, while noting that non-equilibrium dynamics play a crucial role in many systems and are rightly garnering increased attention in the field~\cite{grainger2025equilibrium}.

The terminology used to study biological systems with multiple stable states, often influenced by frameworks from other fields, can lead to multiple definitions of terms, and the concept of stability is a prime example. Following the emphasis of Lewontin, and as is emphasised in mathematical considerations of dynamical systems, we take stability to mean the existence of an invariant set that is a solution or state that is maintained, absent any significant perturbation. With respect to ~\eqref{eq.ODE_model}, local (asymptotic) stability of an equilibrium $\bar{x}$ means that there exists a neighbourhood of $\bar{\mathbf{x}}$ such that for any $\mathbf{x}(0)$ in that neighbourhood, $\lim_{t\to\infty}\mathbf{x}(t) = \bar{\mathbf{x}}$. In contrast, an equilibrium $\bar{\mathbf{x}}$ is globally (asymptotically) stable if $\lim_{t\to\infty}\mathbf{x}(t) = \bar{\mathbf{x}}$ for any initial condition $\mathbf{x}(0)$, and in this case, the system does not admit multiple stable states.

Key to the definition of stability is characterising the class of perturbations to which a system is stable. From a purely empirical standpoint, the absence of a species in a population may be considered stable, as new individuals cannot appear. However, a mathematical or community assembly perspective may categorise this as unstable if the population can be invaded by a small introduction of the absent species. Fully understanding whether this species introduction falls within the class of perturbations is key to defining the stability of the state in the absence of the species. 

\begin{figure}[htp]

\centering
\includegraphics[width=.99\textwidth]{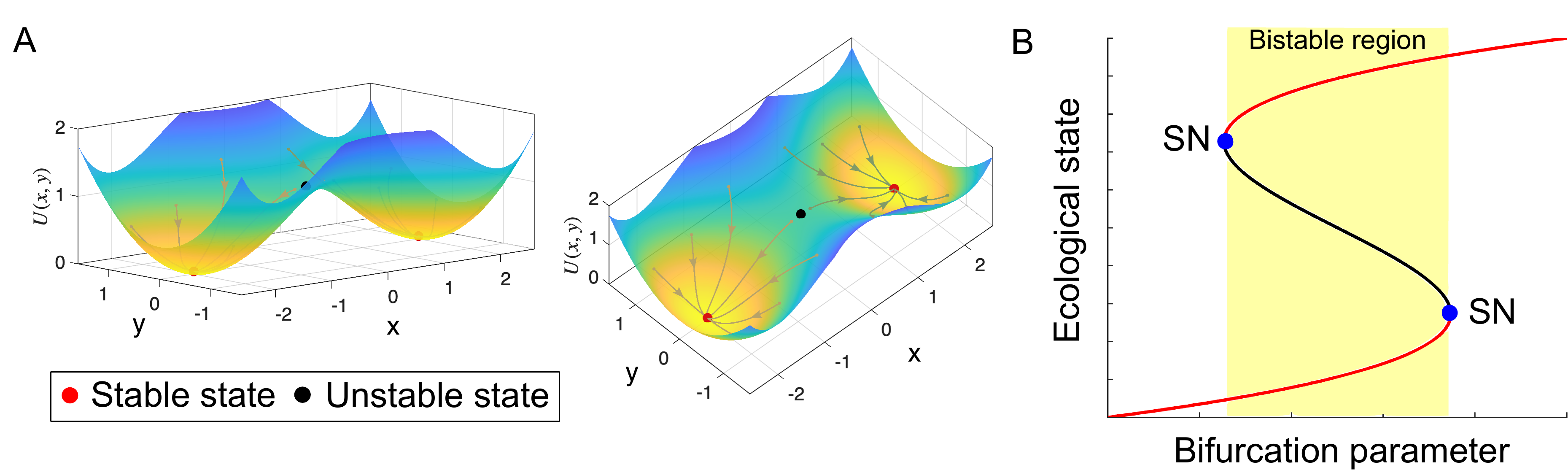}
\caption{\textbf{A}: The potential surface $U(x, y)$ for a bistable system, where stable equilibria correspond to local minima (valleys) and the unstable equilibrium corresponds to a local maximum (ridge) separating the two basins of attraction. The depth and width of each valley reflect the resistance of the corresponding stable state to perturbation. \textbf{B}: One-parameter bifurcation diagram illustrating how bistability arises and disappears as a system parameter is varied. Solid lines indicate stable branches of equilibrium solutions, and dashed lines indicate unstable branches. At each saddle-node bifurcation point (SN), a stable and an unstable equilibrium collide and annihilate, marking the boundary of the bistable region (shaded).}
\label{fig:terms_fig}
\end{figure}

The geometry of basins of attraction, specifically their shape, depth, and extent, plays a crucial role in determining how a system responds to perturbation, and is often understood through the lens of \emph{potential functions}. For an equilibrium solution $\bar{\mathbf{x}}$ of ~\eqref{eq.ODE_model}, the basin of attraction is the set of all initial conditions $\mathbf{x}(0)$ such that $\lim_{t\to\infty}\mathbf{x}(t)=\bar{\mathbf{x}}$. The system ~\eqref{eq.ODE_model} is said to be of gradient-type if there exists a scalar function $U$ such that $\tfrac{d}{dt}\mathbf{x} = f(\mathbf{x}) = - \nabla U(\mathbf{x})$. Trajectories of ~\eqref{eq.ODE_model} will then be decreasing along the potential $U$, so that stable equilibria correspond to local minima of $U$, with unstable equilibria corresponding to local maxima or saddles. This motivates thinking of potential functions as ``stability landscapes'' that allow us to easily visualise basins of attraction for low-dimensional systems; Figure~\ref{fig:terms_fig}A shows the potential landscape for a bistable system. When perturbed or subject to random noise, a system near a stable state with a broad, shallow potential landscape will exhibit very different behaviour than one near a stable state with a narrow but deep potential. Potentials can help predict and quantify responses to perturbations, thereby determining how easily a system can transition between stable states. Since most systems are not exactly of gradient type, there has been significant interest in constructing ``quasi-potentials'' in certain distinguished limits or when ~\eqref{eq.ODE_model} is subject to stochastic forcing~\cite{nolting2016balls}; the stability landscapes of both the forest-savanna and coral-reef models discussed in Section~\ref{sec.previous} have recently been studied via quasi-potential methods~\cite{xu2021unifying,xu2025global}. However, the quasi-potential is generally non-smooth at the boundaries between basins of attraction, and may inherit fractal structure when those boundaries are themselves fractal~\cite{Graham1984,Graham1986,Graham1991}, a more common occurrence in high-dimensional systems. In such cases, the quasi-potential provides reliable intuition about stability within each basin, but quantitative predictions about transition costs near the separatrix should be treated with caution.

While we restrict our discussion to systems of multiple \textit{stable} states---as invariant, stable states have more tractable and well-understood properties---similar conclusions may be drawn for systems that display persistent phases, such as quasi-static or transient systems (see also Section~\ref{sec.discussion}). Empirically, behaviours of these systems may be similar, and given data collection limitations, blurring the sharp distinction between these two concepts may prove most practical. 

Three further concepts are central to understanding the dynamical implications of multiple stable states: bifurcations, tipping, and hysteresis. Bifurcation analysis, discussed in detail below, provides the primary mathematical tool for locating parameter regimes in which multiple stable states coexist. Tipping refers broadly to rapid, disproportionate transitions between states in response to relatively small changes, though the term carries many definitions and is increasingly used loosely outside the scientific literature. Within the mathematical literature, tipping is classified by mechanism: B-tipping (bifurcation-induced), N-tipping (noise-induced), and R-tipping (rate-induced)~\cite{hastings2026tipping}. Hysteresis, the asymmetry between the parameter thresholds for forward and backward transitions, is a defining signature of bistable systems and a central challenge for ecological restoration. Together these concepts characterise how and when transitions between stable states occur, with direct implications for management.

The existence and extent of multiple stable states in a model of the form ~\eqref{eq.ODE_model} can be systematically characterised through bifurcation analysis, as illustrated in Figure \ref{fig:terms_fig}B. As a system bifurcation parameter is varied (representing, for example, environmental conditions), the number and stability of equilibria can change qualitatively at critical parameter values called bifurcation points~\cite{strogatz2001nonlinear}. In systems with multiple stable states, the most common such transition is the saddle-node (or fold) bifurcation, at which a stable and an unstable equilibrium collide and annihilate one another. Pairs of saddle-node bifurcations, labelled SN in Figure \ref{fig:terms_fig}B, delimit a region of parameter space in which two stable equilibria coexist simultaneously, i.e. the bistable region. Identifying whether a mathematical model supports saddle-node bifurcations and locating the bistable parameter range are, therefore, primary tools for assessing whether a model is capable of producing multiple stable states and for predicting under what environmental conditions transitions between states may occur.

We note that the concept of ecological resilience, broadly defined as a system's capacity to absorb disturbance and retain its structure, is closely linked to the ideas in this paper. However, the definition of ecological resilience is itself the subject of a substantial, contested literature beyond our scope here. We hence direct the interested reader to~\cite{holling1973resilience,walker2004resilience} for detailed, foundational treatments.

\begin{table}[!tbp] 
%\begin{tabular}{p{3.5cm}p{9cm}p{3cm}} %{llll} 
\begin{tabular}{>{\raggedright\arraybackslash}p{2.5cm}>{\raggedright\arraybackslash}p{9.5cm}>{\raggedright\arraybackslash}p{2cm}}
\textbf{Term }                             & \textbf{Definition} & \textbf{Ref(s)} \\ \hline
Stable state & An asymptotically attracting community configuration that persists in the absence of significant disturbance over ecologically relevant timescales.    & ~\cite{holling1973resilience}               \\ \hline  
Multiple stable states         & Existence of two or more stable states
                                    under the same set of environmental conditions.  & ~\cite{beisner2003alternative,may1977thresholds} \\ \hline
Ergodicity & The property of a dynamical system whereby time averages converge to spatial averages, so that observing many realisations of a system is equivalent to observing a single realisation over a long time horizon.  & ~\cite{walters2000introduction}              \\ \hline
Basin of attraction & The set of initial conditions whose trajectories converge to a given attractor (e.g. a stable state).  & ~\cite{holling1973resilience} \\ \hline
Separatrix  & The boundary in state space separating the basins of attraction of two or more stable states, which may have a very complex (even fractal) structure in higher dimensions. & ~\cite{strogatz2001nonlinear,freidlin_random_1984} \\ \hline
Potential            & A scalar function $U$ whose values represent the ``height'' of a stability landscape, formally defined for systems of the form $\dot{\mathbf{x}} = -\nabla U(\mathbf{x})$. Stable equilibria correspond to local minima of $U$, and unstable equilibria to local maxima or saddle points. The depth and width of valleys reflect resistance to perturbation. &~\cite{holling1973resilience,may1977thresholds,strogatz2001nonlinear}  \\ \hline
Quasi-potential  & A generalisation of the potential to non-gradient systems with stochastic forcing. The quasi-potential quantifies the minimum cost, in the sense of Freidlin--Wentzell large deviation theory, of a noise-driven transition from one system state to another. & ~\cite{cameron_finding_2012,freidlin_random_1984,nolting2016balls}  \\ \hline
Transient dynamics                & Dynamical behaviour that persists over ecologically relevant timescales but does not represent the long-run asymptotic state of the system. & ~\cite{hastings2004transients,hastings2018transient}    \\ \hline
Alternative transient dynamics   & The existence of two or more transient states
                                    under the same set of environmental conditions.    &~\cite{fukami2011commmunity}    \\ \hline
Hysteresis & Different critical thresholds for forward and backward transitions between states, such that the pathway to return to the original state differs from the pathway of transition away from that state. &~\cite{scheffer2001catastrophic,may1977thresholds} \\ \hline
Bifurcation & A qualitative change in dynamics (e.g., a steady-state appearing/ceasing to exist or gaining/losing stability) as system parameters vary. & ~\cite{strogatz2001nonlinear}  \\ \hline
Tipping point & A critical threshold beyond which a rapid transition to another distinct stable state occurs.  &~\cite{lenton2008tipping,scheffer2009early} \\ \hline
Positive (negative) feedback loop & 
A closed chain of interactions between system components whose net effect either amplifies (positive) or dampens (negative) a perturbation -- see Section~\ref{sec.positive_feedback}. & ~\cite{deangelis2012positive}  \\ \hline
% Endogenous/Exogenous feedback& &
% Basin of attraction
% Conservative systems
% Disturbance
% Stochastic...?
\\
\end{tabular}
\caption{Common terms and concepts related to multiple stable states in ecology.}\label{table:terms}
\end{table} 

%With the key terms suitably defined, we may further explore unifying features of systems with multiple stable states. 

\subsection{The role of positive feedback in multiple stable state theory}\label{sec.positive_feedback}
While it is accepted that positive feedback loops and alternative stable states are linked, a rigorous mathematical exploration of why this is the case is lacking. The term ``positive feedback'' has been used by different authors to refer to numerous different but related effects, often leading to confusion in the literature (see Section~\ref{sec.history}). We give a precise mathematical definition of a positive feedback loop below (refining the informal description from Table~\ref{table:terms}) and illustrate how this concept provides a unifying picture of multiple stable states for a broad class of mathematical models.

Across diverse ecosystems, most theoretical and empirical studies invoking multiple stable states explain them through positive feedback loops, often arising from positive interactions or strong biotic–abiotic feedback~\cite {beisner2003alternative,scheffer2001catastrophic}. For example, according to K\'efi et al.~\cite{kefi2016can}, ``a positive feedback is a necessary (but not sufficient) condition for systems to have alternative stable states,'' a statement which requires some unpacking and context, but which can be made mathematically precise. Following DeAngelis et al.~\cite{deangelis2012positive} and Yodzis~\cite{yodzis1989introduction}, in the context of an equilibrium solution $\bar{\mathbf{x}}$ to ~\eqref{eq.ODE_model}, we define a feedback loop as follows: At an equilibrium $\bar{\mathbf{x}}\in \mathbb{R}^N_+$, the entries of the Jacobian matrix $\mathbf{J}$ of the system are given by
\begin{equation}\label{eq.Jacobian}
(\mathbf{J})_{i,j}(\mathbf{x}) = \frac{\partial}{ \partial x_j}f_i(\mathbf{x}) \vert_{\mathbf{x} = \bar{\mathbf{x}}}, \quad i,j\in\{1,\dots,N\}.
\end{equation}
Intuitively, we can think of the entry $(\mathbf{J})_{i,j}(\bar{\mathbf{x}})$ as the effect that species $j$ has on species $i$ in the vicinity of the equilibrium $\bar{\mathbf{x}}$. There is a \textbf{positive feedback loop} for species $\mathbf{i}$ at $\bar{\mathbf{x}}$ if there exists a non-empty sequence of distinct species $\{i_1,\dots,i_k\}$ such that
\[
L = \text{sgn}\left((\mathbf{J})_{i,i_1}\right) \times \text{sgn}\left((\mathbf{J})_{i_1,i_2}\right) \times \dots \times \text{sgn}\left((\mathbf{J})_{i_k,i}\right) > 0.
\]
We may not have $i_j = i$ for any $j\in\{1,\dots,k\}$, and hence feedback loops must contain at least two species. By ``distinct species'', we also mean that repeated intermediaries are ruled out. If there is a positive feedback loop at an equilibrium $\bar{\mathbf{x}}$, due to the continuity of $\mathbf{f}$, this feedback loop must be present in some neighbourhood of $\bar{\mathbf{x}}$. We also note that the definition above is related to, but distinct from, Levins' definition of feedback at level $k$, which would involve summing up over all feedback loops (as defined above) involving $k$ species with the other $N-k$ species at equilibrium~\cite{levins1974discussion}. Moreover, the definition above makes it clear that a positive feedback loop need not involve purely positive interactions; it can result from a chain of negative interactions or from a chain of both positive and negative interactions.

Mathematically, the statement that positive feedback loops are necessary for alternative stable states was first formalised as Thomas' conjecture, which states that if the system ~\eqref{eq.ODE_model} has at least two non-degenerate equilibria (i.e. $\det(\mathbf{J}(\bar{\mathbf{x}}))\ \neq 0$), then there must exist a positive feedback loop somewhere in the phase space~\cite{thomas1981relation}; Thomas' conjecture was first proven in full generality by Soul\'e in 2003~\cite{soule2003graphic}. Thus, the absence of positive feedback loops in a model can rule out multiple stable states, whereas their presence indicates that further analysis is warranted. For some classes of systems, such as those with monotone structure, specific methods have been developed to detect bistability~\cite{angeli2004detection}. Other fields, notably chemical reaction network theory, have pursued the development of automated tools to identify positive feedbacks and multiple stable states given a set of interactions and assumptions on their functional forms~\cite{craciun2005multiple}.

\begin{figure}[!htbp]
    \centering
    \includegraphics[width=0.85\linewidth]{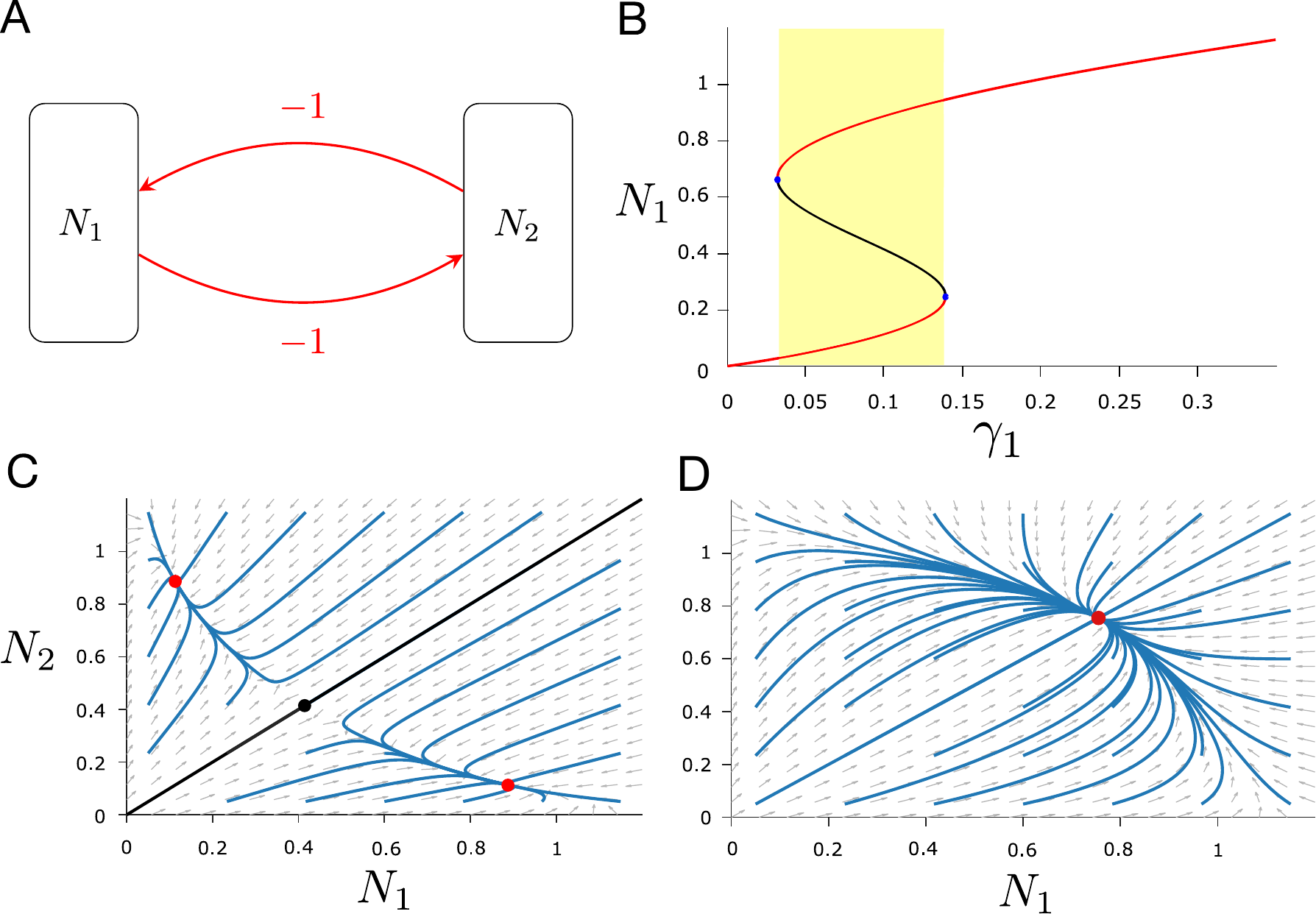}
    \caption{\textbf{A}: Loop diagram illustrating the positive feedback loop composed of two negative interactions in the two-species competition model ~\eqref{eq.LV_model}. Solid red lines indicate stable branches of equilibrium solutions, solid black lines indicate unstable branches of equilibrium solutions, and the filled blue dots indicate the location of saddle-node bifurcation points at the onset/offset of multistability. The bistable parameter range is highlighted in green. \textbf{B}: One-parameter bifurcation diagram as a function of migration rate $\gamma_1$ ($\gamma_2 = 0.1$, $c_1 = c_2 = 2$). \textbf{C}: Phase space plot of the system ~\eqref{eq.LV_model} with $\gamma_1=\gamma_2 = 0.1$ and $c_1 = c_2 = 2$ showing multiple stable states and their basins of attraction. Solid blue lines indicate solution trajectories tending to one of the two alternative stable states (filled red dots). The solid black lines mark the stable manifold of the unstable fixed point (black dot). \textbf{D}: Phase space plot of the system ~\eqref{eq.LV_model} with $\gamma_1=\gamma_2 = 0.1$ and $c_1 = c_2 = 0.5$ showing a single globally stable state.}
    \label{fig:competition}
\end{figure}

For relatively low-dimensional ecological models, so-called ``loop diagrams'' (formally, directed graphs) are a useful tool for representing and quickly identifying positive feedback loops graphically. To form the loop diagram for a given model of the form ~\eqref{eq.ODE_model}, we begin with a set of $N$ vertices or nodes, one corresponding to each species, and then draw an arrow from species $j$ to $i$ with the corresponding sign of ($\mathbf{J})_{i,j}$~\cite{levins1974discussion,deangelis2012positive,yodzis1989introduction}. To illustrate these concepts by way of a concrete example, consider a classical Lotka-Volterra competition model with migration ($\gamma_1,\,\gamma_2>0$), with dynamics given by:
\begin{equation}\label{eq.LV_model}
\begin{aligned}
\frac{d}{dt}N_1 &= N_1 (1 - N_1 - c_1 N_2 ) + \gamma_1 \\
\frac{d}{dt}N_2 &= N_2 (1 - c_2 N_1 - N_2 ) + \gamma_2,
\end{aligned}    
\end{equation}
where $c_1$ and $c_2$ are strictly positive competition parameters. We include the migration terms in ~\eqref{eq.LV_model} to rule out boundary equilibria (on the boundaries of the feasible region $N_1,N_2\geq 0$) and hence restrict to interior equilibria only. The Jacobian for the system ~\eqref{eq.LV_model} is given by
\begin{equation}\label{eq.Jacobian_LV}
\mathbf{J}(N_1,N_2) = 
\begin{bmatrix}
 1 - 2N_1 - c_1 N_2 & -c_1 N_1 \\
 - c_2 N_2 & 1 - c_2 N_1 - 2N_2
\end{bmatrix} \sim 
\begin{bmatrix}
   * & {\color{red} -1} \\
   {\color{red} -1} & *
\end{bmatrix},
\end{equation}
where we use the $\sim$ notation to denote the sign structure of the Jacobian. The loop diagram for the simple competition model ~\eqref{eq.LV_model} is shown in Figure~\ref{fig:competition}A with the arrows representing the interactions between the species and their signs drawn from the Jacobian matrix ~\eqref{eq.Jacobian_LV}. The definition of positive feedback loops requires unique sequences of species and cannot include the species itself. We thus do not need to know the sign of diagonal entries to assess the presence or absence of positive feedback. From the Jacobian above we can identify that there is a feedback loop from species $1$ to species $2$ and vice versa, i.e., $L = \text{sgn}\left((\mathbf{J})_{1,2}\right) \times \text{sgn}\left((\mathbf{J})_{2,1}\right) > 0$ is a positive feedback loop. This positive feedback loop consists solely of negative interactions, and thus the motif that generates it is generally referred to as \emph{mutual inhibition}. Note that we only needed to know that any equilibrium $(\bar{N}_1,\bar{N}_2)$ will have $\bar{N}_1>0$ and $\bar{N}_2>0$, and not the exact values, to confirm the presence of a positive feedback loop in this example. Choosing $c_1 = c_2 = 2$ (strong competition regime), Figure~\ref{fig:competition}B is a one-parameter bifurcation diagram for the system ~\eqref{eq.LV_model} illustrating how alternative stable states emerge via a saddle-node bifurcation as the migration rate $\gamma_1$ is increased and then how the system reverts to a single stable equilibrium again when $\gamma_1$ is sufficiently large; the bistable parameter range is highlighted in green. Figure~\ref{fig:competition}C shows a phase-space perspective on the system dynamics in the bistable regime (green region in Figure~\ref{fig:competition}B) with the two stable states marked by red dots and the unstable equilibrium marked by a black dot. Moreover, in this example, the one-dimensional stable manifold of the unstable equilibrium (solid black line) separates the phase space into two disjoint regions, corresponding to the basins of attraction of the stable states.

To illustrate that positive feedback is merely necessary but not sufficient for multiple stable states, consider the system ~\eqref{eq.LV_model} in the weak competition regime with $c_1 = c_2 = 0.5$, so that interspecific competition is weaker than intraspecific self-regulation. The Jacobian sign structure (and hence the associated loop diagram) is unchanged,  both off-diagonal entries remain negative at any interior equilibrium, and hence the positive feedback loop via mutual inhibition identified in Figure~\ref{fig:competition}A persists. However, with weak competition, the system admits only a single, globally stable coexistence equilibrium for all positive migration rates ($\gamma_1, \, \gamma_2$), and no saddle-node bifurcations occur (see Figure~\ref{fig:competition}D).

\begin{figure}[htbp]
    \centering
    \includegraphics[width=0.9\textwidth]{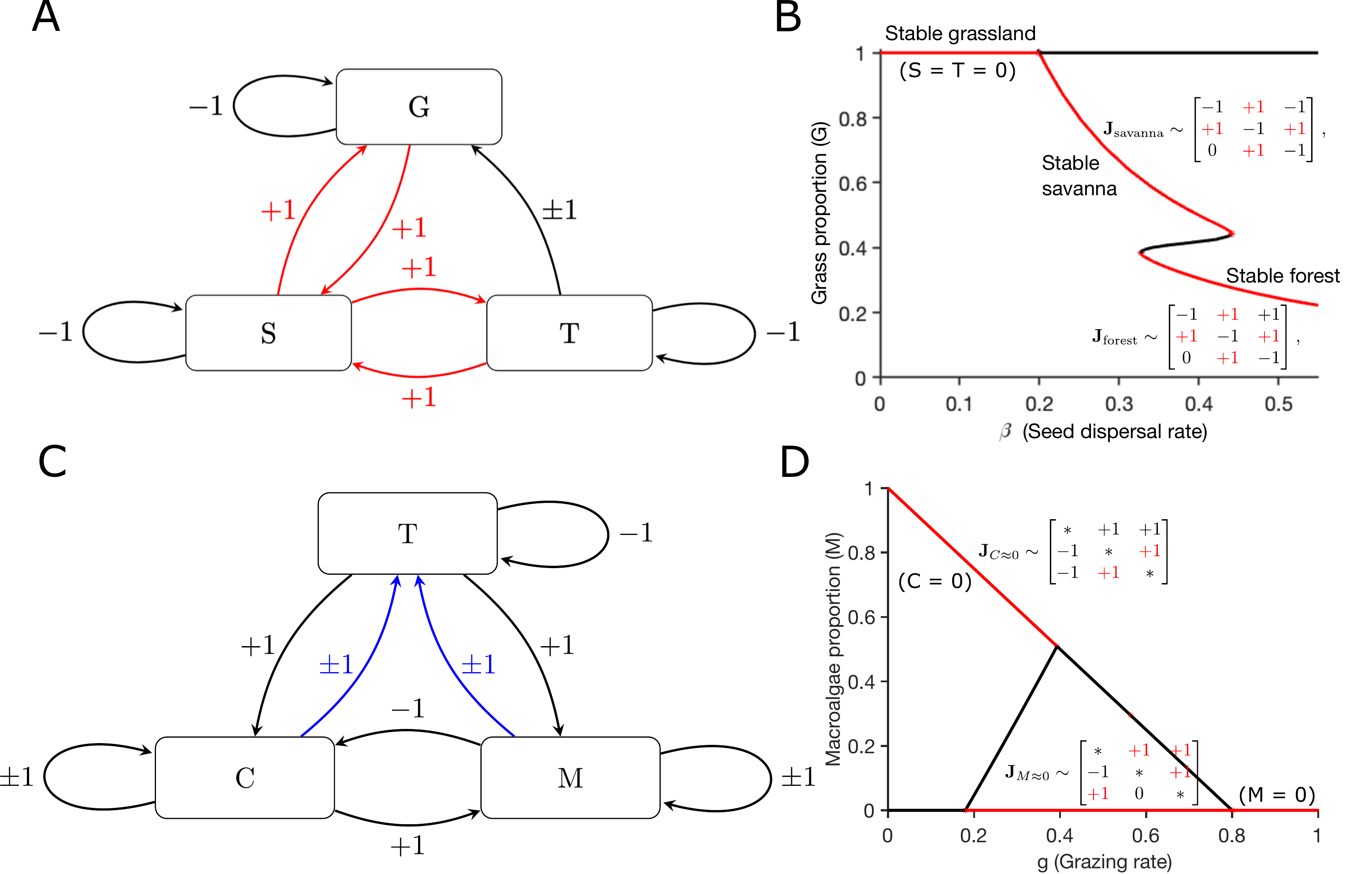}
    \caption{\textbf{A}: Loop diagram for the forest-savanna model ~\eqref{eq.SL_model} illustrating multiple possible positive feedback loops in red, assuming we are away from the grassland equilibrium ($G = 1$, $T=S=0$); the $\pm 1$ value indicates that the sign of the corresponding Jacobian entry can change depending on the equilibrium value. \textbf{B}: Bifurcation diagram for the forest-savanna model ~\eqref{eq.SL_model} varying the rate of savanna tree seed dispersal ($\beta$), illustrating the Jacobian sign structures and positive feedback loops at each alternative stable state. \textbf{C}: Loop diagram for the coral reef model due to~\cite{mumby2007thresholds}, assuming the Jacobian is evaluated away from boundary equilibria; blue arrows highlight possible positive feedback loops that depend on the equilibrium in question. \textbf{D}: Bifurcation diagram for the coral reef model with the Jacobian sign structures evaluated in the neighbourhood of each (boundary) equilibrium restricted to the interior of the admissible phase space with entries generating positive feedback loops in red. Further details on both models and parameter values can be found in Appendix~\ref{sec.models_appendix}.}
    \label{fig.loops}
\end{figure}

For a more complex example, consider the following forest-savanna model~\cite{staver2011tree} with three ``species'' (functional types) of vegetation: grass ($G$), savanna saplings ($S$), and savanna trees ($T$). This model considers the proportion of a landscape that is filled by each of these species, and hence we require $G+S+T =1$ and have dynamics given by:
\begin{align}\label{eq.SL_model}
\begin{cases}
\dot{G} &= \mu S + \nu T - \beta G T, \\
\dot{S} &= \beta G T - \omega(G)S - \mu S, \\
\dot{T} &= \omega(G)S - \nu T,
\end{cases}
\end{align}
where $\mu$, $\nu$ and $\beta$ are all positive parameters, and $\omega:[0,1] \mapsto [0,1]$ models the impact of fire effects on sapling maturation and is a smooth approximation to a shifted Heaviside function with a sharp threshold around $40\%$ grass cover (see Appendix~\ref{sec.savanna}). This highly nonlinear functional response is a phenomenological model of fire as a percolation process~\cite{staver2011tree}. Further details of the model and parameter values used can be found in Appendix~\ref{sec.models_appendix}. The Jacobian of ~\eqref{eq.SL_model} is given by
\begin{align*}
\mathbf{J}(G,S,T) &= 
\begin{bmatrix}
-\beta T & \mu & \nu - \beta G \\
\beta T - \omega'(G)S & -\omega(G)-\mu & \beta G \\
\omega'(G)S & \omega(G) & -\nu
\end{bmatrix} \\
&\approx
\begin{bmatrix}
-\beta T & \mu & \nu - \beta G \\
\beta T & -\omega(G)-\mu & \beta G \\
0 & \omega(G) & -\nu
\end{bmatrix} \sim 
\begin{bmatrix}
* & {\color{red}+1} & {\color{blue} \pm 1} \\
{\color{red}+1} & * & {\color{red}+1} \\
0 & {\color{red}+1} & *
\end{bmatrix},
\end{align*}
where we use the fact that $\omega$'s derivative is zero away from the threshold region around $40\%$ grass cover in the approximation. For this analysis, we assume that equilibria are interior, i.e. $\bar{G},\,\bar{S},\,\bar{T} \in (0,1)$. Moreover, the $\pm$ sign on the arrow from $T$ to $G$ arises owing to the fact that the sign of $\nu-\beta G$ depends on the value of the steady-state. The loop diagram for this model is shown in Figure~\ref{fig.loops}A and immediately highlights positive feedback loops from grass to grass and from saplings to saplings (the interactions comprising these loops are highlighted in red). It is also possible to have a positive feedback loop from trees to trees if grass is sufficiently low ($\bar{G} \in (0,\nu/\beta)$). Figure~\ref{fig.loops}B illustrates bistability between a stable savanna equilibrium and a stable forest equilibrium (high tree cover) via a bifurcation diagram that allows the seed dispersal rate of trees ($\beta$) to vary; we also highlight the sign structure of the Jacobian along each branch of stable equilibrium solutions.

Figure~\ref{fig.loops}C shows the loop diagram for the coral reef model of Mumby et al.~\cite{mumby2007thresholds}, which contains as state variables the proportions of turf, coral and macroalgae that cover a reef. In this case, Jacobian analysis alone is ambiguous, and we use ecologically plausible parameters in order to determine the equilibrium values of signs in the Jacobian. We observe in Figure~\ref{fig.loops}C that the presence or absence of positive feedback loops depends on the signs of the effect of coral on turf and macroalgae on turf; the sign of these effects depends on exactly where the Jacobian is evaluated. This model also features a number of boundary equilibria (i.e. at least one cover type equal to zero) and, as the one-parameter bifurcation diagram in Figure~\ref{fig.loops}D illustrates, the model supports bistability between an equilibrium dominated by macroalgae (with zero coral) and a coral-dominated equilibrium (with no macroalgae present). As the Jacobians overlaid on the stable equilibrium branches in Figure~\ref{fig.loops}D show, there is a positive feedback loop from each species to itself when coral is sufficiently low (e.g. $M \to C \to T \to M$, where $L = -1 \times -1\times +1 = +1$). Similarly, when $M\approx 0$, there is a positive feedback loop involving all three species ($L = +1 \times +1\times +1 = +1$), as well as a positive feedback loop involving just turf and macroalgae arising from mutually positive interactions.

Multiple stable states can also emerge in more complex models as system complexity and the number of interacting species increase. Positive feedback loops remain a necessary condition for the existence of multiple stable states as system complexity increases, though the mechanisms through which they arise may change and are less amenable to straightforward ecological interpretation. Recent work by Aguadé-Gorgorió et al.~\cite{aguagegorgorio2024taxonomy} on species-rich community models reveals that increasing complexity can generate new multistability regimes with no direct analogue in low-dimensional systems. They identify four such regimes, namely global bistability, local multistability, mutual exclusion, and cliques, whose character depends on the sign and heterogeneity of interspecific interactions rather than on bistability at the level of individual species. Similarly, Bunin showed that in large communities with randomly sampled competitive interactions, a transition to multistability occurs with sufficiently strong and heterogeneous interactions~\cite{bunin2017ecological}. Crucially, in these much higher-dimensional models, even the regimes that emerge from purely competitive dynamics without any species-level positive interaction must contain positive feedback loops in their Jacobian sign structure, possibly arising through chains of negative interactions, as in the mutual inhibition motif of our Lotka-Volterra example.

\subsection{Common mechanisms generating positive feedback in ecological systems} 

Here, we present examples of mathematical and ecological mechanisms that commonly generate positive feedback loops and potential multistability in ecological models. These mechanisms are not exhaustive, and our categories of mechanisms are not mutually exclusive; in many systems, multiple mechanisms may be present and interact with one another.

\subsubsection{Strongly nonlinear functional responses} A nonlinear response in at least some state variables is a trivial mathematical requirement for the \textit{existence} of alternative states; thus, linear models can have at most one steady state. Nonlinearity is required to have all species present at multiple interior equilibria (i.e. no presence/absence equilibria with only a subset of the species at nonzero levels)~\cite{petraitis2013multiple}. Nonlinear functional responses include highly nonlinear responses, like sigmoidal functions, non-monotonic functional responses, and positive density dependence, i.e., the Allee effect, and are often categorised in terms of Holling's classic three functional response types with a fourth sometimes added to include non-monotone or ``dome-shaped'' response curves~\cite{holling1959components}. There is a vast literature on alternative stable states in predator-prey and consumer-resource models with strongly nonlinear functional responses. For example, May showed that type II and type III functional responses can generate alternative stable states in consumer-resource models~\cite{may1977thresholds} . 
%Beisner et al. demonstrate the effects of Hill functions~\cite{beisner2003alternative}.

\subsubsection{Path-dependent assembly} This category covers mechanisms like priority effects, ecosystem engineers, and historical contingency effects, i.e., the order in which species arrive and then tailor the environment to themselves ``locks in'' one stable state, but if the order of arrival were different, the outcome would be different~\cite{fukami2015historical}. However, this should still be part of positive feedback loops, i.e., tailoring the environment to reinforce the state that favours the occupying community is itself a positive feedback loop. Part of the reason this is usually distinguished as a distinct mechanism in the literature is that it explicitly requires biotic-abiotic interactions, unlike models with only species-species interactions.

\subsubsection{Biotic-abiotic feedbacks and ecosystem engineering} A third important class of mechanisms involves feedbacks between organisms and their abiotic environment, in which a community modifies physical or chemical conditions in ways that reinforce its own persistence~\cite{wilson1992positive, cuddington2009ecosystem}. In shallow lakes, aquatic macrophytes stabilise water clarity by reducing sediment resuspension and competing with phytoplankton for nutrients, thereby reinforcing the clear-water state; this biotic-abiotic coupling is precisely the mechanism underlying the bistability formalised by Scheffer~\cite{scheffer1989alternative}. In the forest-savanna system, the suppression of the grass layer by closed-canopy forest reduces fire frequency, which in turn favours tree recruitment and further canopy closure; this positive feedback loop is captured in the highly nonlinear fire function $\omega(G)$ in model ~\eqref{eq.SL_model}. Unlike path-dependent assembly, these feedbacks are ongoing and self-reinforcing at equilibrium rather than being contingent on historical arrival order, and their disruption is often a primary target of restoration interventions~\cite{suding2004alternative}.

\subsubsection{Mutualism and facilitation} Positive interspecific interactions, such as mutualism and facilitation, can also generate positive feedback loops and hence multiple stable states, particularly in stressful environments where facilitation between species is strong relative to competition. For example, in dryland ecosystems, established vegetation facilitates the survival of neighbouring plants by reducing soil erosion and ameliorating microclimatic conditions, creating a positive feedback that maintains the vegetated state. Below a threshold of vegetation cover, facilitation collapses and the system tips to a bare-soil state~\cite{kefi2016can, pichon2024interplay}. These facilitative interactions generate positive off-diagonal Jacobian entries that contribute to the positive feedback loops required for multistability.

\section{Discussion} \label{sec.discussion} %(2500 words)
%4 Main Ideas: feedback loops, stability and tipping points, space for time, transients
%Conclusions from the historical section--definitional clarity is important; there is a long dating thread of feedback as a topic of interest; the mathematical framework can be useful to better provide coherence to examples and look for commonalities; there is still a lot of debate around classifying these systems though.
Multiple stable states are now well established as a central framework in theoretical and applied ecology, yet a unified account of the mechanisms and structural features common to systems that exhibit them has been lacking. A recurring theme across the historical, empirical, and theoretical perspectives surveyed in Section \ref{sec.previous} is that definitional clarity matters: precisely specifying key features, such as stability, ecological states, perturbations, and feedbacks, allows for a more productive conversation on identifying systems of multiple stable states. The mathematical framework introduced in Section \ref{sec.math} provides exactly this specificity, and in doing so exposes a unifying picture: positive feedback loops, formalised through the Jacobian sign structure, are a necessary feature of systems with multiple stable states, and their presence can be identified and compared across ecologically diverse systems using the loop diagram approach. Thomas' conjecture~\cite{soule2003graphic} establishes that positive feedback loops are necessary for systems of the form \eqref{eq.ODE_model} to admit multiple stable equilibria, and the loop diagram framework then translates this mathematical requirement into a concrete, interpretable tool. Applying this framework to the Lotka-Volterra competition, forest-savanna, and coral reef models reveals that superficially different ecological systems share a common mathematical signature, even when the biological mechanisms generating it differ substantially.

The scale at which to analyse system feedback and the ecosystem in general may require careful consideration. One challenge to identifying or characterising the feedbacks is that even within one system, different feedback loops, and their resulting effects, may occur on different scales, spatially or temporally~\cite{levin1992problem, kefi2016can}. Key to Yodzis's definition of feedback is an understanding of the timescales on which the feedbacks occur~\cite{yodzis1989introduction}. However, there may be spatial and temporal mechanisms interacting across a range of scales, such as patterns trickling up from smaller to large scales~\cite{levin1992problem, suding2009threshold}. For example, the feedback loops that lead to self-organising species begin with individual responses~\cite{kefi2016can}. Species that engineer their environments to further their survival create a lasting physical impact, introducing a temporal lag in ecosystem recovery (a scale change in time)~\cite{cuddington2009ecosystem}. More broadly, selecting a temporal or spatial scale at which to study a system must be governed by the ecological scale of the community of interest, to capture key interactions and community composition turnover or stabilisation,
%try to capture all interactions of the key player--be as broad as needed but no more; time scale based on species lifespan, spatial scale based on individual spatial impact
although these choices are likely influenced by practical constraints, such as associated research costs or time constraints. Here, improvements in data availability driven by technological advances, such as increased spatial scale or density of data, may ameliorate some observation constraints. Technologically advanced techniques, such as satellite or drone remote sensing, automated data processing methods, and even mobile applications, can complement traditional data collection techniques to provide a fuller understanding of ecosystem dynamics~\cite{manfreda2018use,marvin2016integrating}. Additionally, integrating multiple sources of data may provide the most insight into phenomena across scales, while also possibly reducing the effect of data bias~\cite{zipkin2021addressing}. Harnessing these advances to study systems with multiple stable states will likely reveal a clearer picture of the system's spatial diversity and the patterns that emerge. With long-term temporal studies still presenting major challenges, the space-for-time substitution may prove vital for harnessing these data-collection advancements.

In restoration and management contexts, a better understanding of commonalities and drivers behind systems of multiple stable states may also have promising implications. Dedicating time to identifying the feedbacks present, the factors that control them, the scales at which they occur, and their strength may play key roles in a full understanding of the underlying dynamics. It may be most advantageous to target these feedbacks specifically in restoration. For example, in coral reefs, rebuilding depleted grazing fish populations to reinforce the grazing feedback and disrupt macroalgae growth may be complementary to coral outplanting~\cite{sha2026alternative}. If ecosystem engineers are present in a system, it may be most practical to focus restoration resources on this population, to enhance the system feedback loops~\cite{byers2006using}. For abiotic restoration, if harsh environmental conditions or disturbance regimes play a major role in dynamics, they may be crucial to maintaining the desired state, and without restoration of this abiotic landscape, ecosystems may be resistant to state changes~\cite{didham2005systems}. It may also be favourable to partition resources between biotic and abiotic players in feedback loops to optimise efficiency~\cite{byers2006using}.

In a theoretical context, creating representative models of ecosystems is key to creating insightful conclusions that can be used to understand, predict, or manage ecological systems. This review underscores the importance of codifying interactions and feedback in creating a representative model. Through the loop diagram structure and analysis of the Jacobian, the presence and influences of positive feedback loops emerge more clearly, and while not sufficient proof of multistability, this analysis may provide support for conclusions about the role of feedback in an ecological system. Foundational to theoretical conclusions is a clear definition of stability in the analysis of an ecosystem. Clearly defining the class of perturbations with which one defines stability with respect to is key to better understanding. The question may be asked, however, how often the choice of stability in a system's model is one of convenience, or if a system could be similarly modelled in a transients-focused framework. Consider, for example, an ecological system experiencing frequent oscillatory environmental perturbation~\cite{mushet2020alternative}. Would stable or transient states be more appropriate to describe these communities? Could we instead consider how to lessen the divide between stable and persistent frameworks?
 
%\subsection{Implications for systems of multiple transient states} \label{subsec:persistent}
While there has been debate over the existence of definitionally-strict multiple stable states, differentiating between stable and long-term transient dynamics might be inhibiting our understanding of these types of systems. Stochasticity may disrupt how system behaviour is observed, and can make determining the stability of a state difficult to distinguish empirically~\cite{abbott2017alternative}. Stochasticity can also create the appearance of multiple stable states or obscure the differentiation of multiple stable states~\cite{abbott2017alternative, abbott2021mapping}. Additionally, while the space-for-time technique may still prove useful in the case of multiple transient states, it is not without further complication; the observed length of a transient state may not correlate directly with the underlying stationary distribution~\cite{morozov2020long,morozov2024longliving}.

From a modelling perspective, we must often choose to model ecosystems as either having persistent transients or multiple stable states, and it is important to consider how this choice may impact our understanding of a system. For example, nitrogen, phytoplankton, and zooplankton (NPZ) systems undergo blooms and die-offs, transitions between impermanent system compositions. System feedback loops show similarities in structure to many of the multiple stable state examples given above, yet the system is not typically modelled with multiple stable states~\cite{franks2002npz}. This system could be reframed in a model with multiple stable states, but it remains open as to whether this reframing would capitalise on similarities in a practically useful way. More generally, if transient dynamics can be reframed as asymptotic dynamics with multiple stable states, or vice versa, when is it advantageous to do so?

In the context of management actions, the differentiation of transients and stable state frameworks may prove important. Indeed, unlike stable states, truly transient states need not require intervention to return to a previous state; however, in the case of long-term transients, the temporal scale of persistence of a state may still instigate management action. As transients and stable states may be difficult to empirically differentiate, yet can respond differently to disturbance or restoration forces, considering the implications of management actions in both a transient and stable framework may prove to be the most informative of the full range of possible outcomes~\cite{boettiger2021ecological}.
%A system showing limited response to intervention or disruption may not necessarily be buffered from tipping.

%\section{Future Directions and Challenges}
%The challenge lies in formalising the analysis of natural phenomena--translating key features from observed behaviour into data and mathematical models. Are we capturing all of the patterns or features in the system of interest? Are we translating these appropriately in to quantitative data and models? 
Despite significant recent progress, there remain many important open questions regarding multiple stable states and the ecosystems that exhibit them. At a high level, the challenge lies in determining if we are capturing and focusing on relevant patterns or features in systems of multiple stable states and if we are appropriately translating these patterns into quantitative data and models. The question still remains whether there are characteristic features that could be used to better identify multistable ecological systems. To this end, are there more observable features in a system that can strengthen the conclusion of multiple stable states? With what frequency do these occur? In quantifying the features we can capture, questions also arise about which theoretical framework is most effective for a given ecosystem. This decision plays out in defining problem scope, terms, and model setup. On the theoretical side, extending approaches like the loop diagram to higher-dimensional and spatially explicit models, where Jacobian sign structure alone may be insufficient to determine multistability, remains an important challenge. The relationship between multiple stable states and persistent transient dynamics also deserves deeper investigation. As we have argued, the two phenomena may be empirically indistinguishable, yet they carry different implications for management and restoration. A unified framework of the kind we have outlined here is only valuable if it informs practice: the most important open question may be how best to translate the mathematical insights and perspectives on multiple stable states into actionable guidance for ecosystem management and restoration.

\section*{Statements and Declarations}

\subsection*{Acknowledgements} 

JP was supported by the U.S. Department of Energy, Office of Science, Office of Advanced Scientific Computing Research, Department of Energy Computational Science Graduate Fellowship under Award Number DE-SC0024386. AH was supported in part by the US NSF (grant no. 2025235).\\

\noindent We thank Madeline Jarvis-Cross and Matthew Neils for thoughtful comments on an earlier draft of this manuscript.

\subsection*{Data Availability} Codes used for model analysis and plot generation are publicly available via an associated GitHub repository: \href{https://github.com/patterd2/multiple_stable_states}{github.com/patterd2/multiple\_stable\_states}.

\subsection*{Competing Interests} The authors have no competing interests to declare.

\subsection*{Disclaimer}

Neither the United States Government nor any agency thereof, nor any of their employees, makes any warranty, express or implied, or assumes any legal liability or responsibility for the accuracy, completeness, or usefulness of any information, apparatus, product, or process disclosed, or represents that its use would not infringe privately owned rights. Reference herein to any specific commercial product, process, or service by trade name, trademark, manufacturer, or otherwise does not necessarily constitute or imply its endorsement, recommendation, or favouring by the United States Government or any agency thereof. The views and opinions of authors expressed herein do not necessarily state or reflect those of the United States Government or any agency thereof.

\noindent

\begin{appendices}

\section{Details of models and parameters}\label{sec.models_appendix}

\subsection{Forest-savanna model analysis}\label{sec.savanna}
The dynamics of the three-species forest-savanna model variant are as follows~\cite{staver2012integrating}: 
\begin{align*}
\begin{cases}
1 &= G + S + T , \\
\dot{G} &= \mu S + \nu T - \beta G T, \\
\dot{S} &= \beta G T - \omega(G)S - \mu S, \\
\dot{T} &= \omega(G)S - \nu T.
\end{cases}
\end{align*}

\begin{table}[!htbp]
\centering
\begin{tabular}{lll}
\hline
\textbf{Symbol} & \textbf{Ecological interpretation} & \textbf{Default} \\
\hline
$\beta$   & Savanna sapling birth rate                            & 0.38 \\
$\mu$     & Savanna sapling mortality rate                        & 0.2  \\
$\nu$     & Adult savanna tree mortality rate                     & 0.1  \\
$\omega_{0}$ & Savanna sapling-to-adult recruitment               & 0.9  \\
           & rate at low grass cover                                     &      \\
$\omega_{1}$ & Savanna sapling-to-adult recruitment               & 0.2  \\
           & rate highest grass cover                           &      \\
$\theta_{1}$ & Grass cover threshold                            & 0.4  \\
$s_{1}$  & Slope of the sigmoid                                  & 0.01 \\
\hline
\end{tabular}
\end{table}
The sigmoid function describing the recruitment rate of saplings to mature savanna trees as a function of grass cover is given by
\[
\omega(G) = \omega_0 + \frac{\omega_1 - \omega_0}{1 + \exp\left(-(G-\theta_1)/s_1 \right)}, \quad G \in [0,1],
\]
reflecting delayed maturation due to top killing of saplings by fire.

\subsection{Coral Reef Model}\label{sec.coral}
The dynamics of the coral reef model we considered are given by~\cite{mumby2007thresholds}:
\begin{align*}
\begin{cases}
1 &= C + M + T, \\
\dot{M} &= a M C - \frac{g}{M + T} M + \gamma M T, \\
\dot{C} &= r T C - a M C - d C, \\
\dot{T} &= \frac{g}{M + T}M - \gamma M T - r T C + d C.
\end{cases}
\end{align*}
The Jacobian for the coral reef model is given by:
\[
\mathbf{J}(M,C,T) = 
\begin{bmatrix}
a C - \frac{gT}{(M+T)^2} + \gamma T & aM  & \frac{gM}{(M+T)^2} + \gamma M \\
- a C & rT - a M - d & r C \\
{\color{blue}\frac{gT}{(M+T)^2} - \gamma T} & {\color{blue}- r T + d} & - \frac{g M}{(M+T)^2} - \gamma M - rC
\end{bmatrix},
\]
Considering the Jacobian with $C\approx 0$ but positive means we look at $M\approx 1 -g/\gamma$ and $T\approx g/\gamma$ with $M+T\approx 1$, i.e.
\[
\mathbf{J}(M,C\approx 0,T) = 
\begin{bmatrix}
* & aM  & (g + \gamma) M \\
- a C & * & r C \\
(g-\gamma)T & -rT +d & *
\end{bmatrix}.
\]
Assume $g<\gamma$ to determine signs, as this is required for stability of the branch for our chosen parameter set.
\begin{table}[!htbp]
\centering
\begin{tabular}{lll}
\hline
\textbf{Symbol} & \textbf{Ecological interpretation} & \textbf{range/value} \\
\hline
$a$ & rate at which corals are overgrown by macroalgae (yr$^{-1}$) & 0.1 \\
$\gamma$ & rate at which macroalgae spread over algal turfs (yr$^{-1}$) & 0.8 \\
$r$ & rate at which corals recruit and overgrow algal turfs (yr$^{-1}$) & 1.0 \\
$d$ & mortality rate of corals (yr$^{-1}$) & 0.44 \\
$g$ & rate at which herbivores consume macroalgae (yr$^{-1}$) & $(0,\gamma)$ \\
\hline
\end{tabular}
\end{table}

\end{appendices}

\bibliography{refs}% common bib file
\bibliographystyle{abbrv}

\end{document}